\documentclass[doublecol]{epl2}

\title{Efficient $k$-separability criteria for mixed multipartite quantum states}
\shorttitle{Efficient $k$-separability criteria for mixed multipartite quantum states} 

\author{Ting Gao\inst{1}\thanks{E-mail: \email{gaoting@hebtu.edu.cn}} \and Yan Hong\inst{1} \and Yao Lu\inst{2}  \and Fengli Yan \inst{3}\thanks{E-mail: \email{flyan@hebtu.edu.cn}}}
\shortauthor{T. Gao \etal}

\institute{
  \inst{1}  College of Mathematics and Information Science, Hebei
Normal University, Shijiazhuang 050024, China\\
  \inst{2} State Key Laboratory of Low-dimensional Quantum Physics and Department of Physics, Tsinghua University, Beijing 100084,
China\\
  \inst{3}  College of Physics Science and Information Engineering, Hebei Normal University, Shijiazhuang 050024,  China }


\abstract{We investigate classification and detection of entanglement of multipartite quantum states in a very general setting, and  obtain efficient $k$-separability criteria for mixed multipartite states in arbitrary dimensional quantum systems. These criteria can be used to distinguish $n-1$ different classes of multipartite inseparable states and can detect many important multipartite entangled states  such as GHZ states, W states, anti W states, and mixtures thereof. They detect $k$-nonseparable $n$-partite quantum states which have previously not been identified. Here $k=2,3,\cdots,n$. No optimization or eigenvalue
evaluation is needed, and our criteria can be evaluated by simple computations involving
 components of the density matrix. Most importantly, they can be implemented in today's
 experiments by using at most $\mathcal{O}(n^2)$ local measurements. }

\begin{document}

\maketitle

\section{Introduction}

Quantum entanglement is a new quantum resource for tasks that cannot
be performed by means of classical resources,  and has widely been
applied to quantum communication \cite{Ekert91,
BBCJPWteleportation93, BBM92, LoChau, YZEPJB05, GYJPA05,
WangPhysRep,YanGaoEricPRA2011} and quantum computation \cite{RB,BD}. So, it has
theoretical and practical value to study the separability of quantum
states.

There is a substantial bulk of work for bipartite systems, in
particular for the case of qubits. Many well-known (necessary)
separability criteria have been proposed to distinguish separable
from entangled states \cite{Bell, PeresPPT, HorodeckiPPT, HorodeckiReduction, CAprl, HorodeckiRange, NKmajority, Rrealignment, CWrealignment03, CWrealignment02, ACFgeneralrealignment, GHGEcovariancematrix07, GGHEcovariancematrix08}.
All these criteria work very well
in many cases, but are not perfect \cite{HorodeckisRMP2009}.
The separability problem for multipartite and high dimensional systems is much more complicated, since many different classes of entanglement not only under deterministic local operations
and classical communication but even under their stochastic analog occur \cite{BPRST2001,AFOV2008}.  Owing to the complicated structure of
multipartite entangled states, it is difficult to decide to which
class a given multipartite state belongs.

$k$-nonseparable $n$-partite state and $k$-partite entangled $n$-partite state are different concepts, although 2-nonseparable $n$-partite state is the same as $n$-partite entangled $n$-partite state. Both $k$-nonseparability and $k$-partite entanglement can be used to characterize  multipartite entanglement.
As genuine multipartite entanglement and  $k$-nonseparability
are being more and more understood, tools for their detection are
beginning to be developed
\cite{GuhneToth2009,GuhneSeevinck2010,GHGcovariancematrix10, MarcusPRL2010,A.Gabriel2010,GaoHong2010,GaoHong2011}.
Separability criteria for two types of multipartite qubit states were presented in Ref.\cite{GuhneSeevinck2010}.
Gittsovich \textit{et al.} \cite{GHGcovariancematrix10} derived a multipartite covariance matrix criterion detecting multiparticle states that are not fully separable.
This criterion allows detection of bound entangled states which are not detected by other commonly used criteria. However,
some strongly entangled pure states such as the Greenberger-Horne-Zeilinger (GHZ) states are not detected by the multipartite covariance matrix criterion \cite{GHGcovariancematrix10}.
Separability criteria to identify genuinely multipartite-entangled mixed quantum states in arbitrary dimensional systems were provided in Ref.\cite{MarcusPRL2010}.  In Refs.\cite{GaoHong2010,GaoHong2011},
we  obtained  separability
criteria to detect genuinely entangled and nonseparable $n$-partite mixed
quantum states in arbitrary dimensional systems. There were experiments and theory on detecting the various different forms of multi-partite entanglement in noisy W states \cite{PappEnkScience2009, ChoiEnkNature2010, LougovskiEnkNJP2009}. By using $k$-partite entanglement, which is different from $k$-nonseparability,  hierarchies of separability criteria that identify $k$-partite entanglement were given in  \cite{LeviMintertPRL2013}.

$k$-separability provides a fine graduation of states according to their degrees of separability \cite{MKBconcurrencePRL05}.
The detection of genuine $k$-nonseparability in
mixed states is essential. A necessary criterion for $k$-separability  was derived in Ref.\cite{A.Gabriel2010}. However, it needs $\mathcal{O}(2^n)$ local observables to be  implemented in experiment.  In this paper, we aim to present  powerful criteria to identify $k$-nonseparable $n$-partite mixed states in arbitrary
dimensional quantum systems.
 These criteria have the following advantages: First, they can be used to distinguish the different classes of multipartite inseparable states. Second, they can detect many important multipartite entanglement states  such as GHZ states, W states, anti W states, and mixtures thereof. Third, they allow us to detect some classes of $k$-nonseparable $n$-partite quantum states that can not detected by all previously studied criteria. Last but not least, these criteria can be used in today's experiment in an efficient way as they require  at most $\frac{5(n^2-n)}{2}+n+1$ local measurements to be implemented,  at the same time, they are  easy computable, no optimization  process is needed.

\section{Definitions}

An $n$-partite system is described by a Hilbert space $\mathcal{H}$ that decomposes into a direct product of $n$ subspaces $\mathcal{H}=\mathcal{H}_1\otimes
\mathcal{H}_2\otimes\cdots\otimes\mathcal{H}_n$, where the dimension of the Hilbert space $\mathcal{H}_l$ will be denoted by $d_l$.
An $n$-partite pure state $|\psi\rangle\in \mathcal{H}$  is called $k$-separable \cite{MKBconcurrencePRL05, TG2006APB, ABLSclassificationPRL2001}  if there is a
$k$-partition $A_1|A_2|\cdots|A_k$ (a partition of $\{1,2,\cdots, n\}$
 into $k$ pairwise disjoint subsets: $\{1,2,\cdots, n\}=\bigcup \limits_{l=1}^kA_l$ with $A_l=\{j_1^l,j_2^l,\cdots, j_{m_l}^l\}$) such that
\begin{equation}\label{}
|\psi\rangle=|\psi_1\rangle_{A_1}|\psi_2\rangle_{A_2}\cdots|\psi_k\rangle_{A_k},
\end{equation}
where $|\psi_i\rangle_{A_i}$ is the state of
particles $j_1^i$, $j_2^i$, $\cdots$, $j_{m_i}^i$. That is, an $n$-partite pure state is $k$-separable, iff it can be written as a product of $k$ substates.
An $n$-partite mixed state $\rho$ is called $k$-separable if it can be
written as a convex combination of $k$-separable pure states
\begin{equation}\label{}
 \rho=\sum\limits_{m}p_m|\psi_m\rangle \langle\psi_m|,
\end{equation}
where $|\psi_m\rangle$ might be $k$-separable under different
partitions. That is, an $n$-partite mixed state $\rho$ is $k$-separable, iff it has a decomposition into
 $k$-separable pure states. The individual pure states composing a $k$-separable mixed state may be $k$-separable under different partitions.
 $k$-separability provides a fine graduation of $n$-partite quantum states according to their different degrees of separability \cite{MKBconcurrencePRL05}.
In particular, an $n$-partite state is called fully
separable, iff it is $n$-separable. It is called genuinely
$n$-partite entangled, iff it is not bi-separable (2-separable). In
general, $k$-separable mixed states are not separable with regard to any
specific partition, which makes $k$-separability rather difficult to
detect.

There exist  bi-separable  states  that are   entangled with respect to a fixed bipartition. The following states, being  mixtures of states that are separable with respect to some bipartition, still carry some entanglement, i.e. neither it can be written as a mixture of separable states with respect to some fixed bipartition nor as a mixture of fully separable states.
Three qubit states $|\psi_1\rangle=\frac{1}{\sqrt{2}}(|00\rangle+|11\rangle)_{23}|0\rangle_1$, $|\psi_2\rangle=\frac{1}{\sqrt{2}}(|00\rangle+|11\rangle)_{13}|0\rangle_2$, and $|\psi_3\rangle=\frac{1}{\sqrt{2}}(|00\rangle+|11\rangle)_{12}|0\rangle_3$ are 2-separable under 2-partition $1|23$, $13|2$, and $12|3$, respectively.  Their convex combinations   $\rho=p_1|\psi_1\rangle\langle\psi_1|+p_2|\psi_2\rangle\langle\psi_2|+p_3|\psi_3\rangle\langle\psi_3|$ ($p_i>0$, $\sum\limits_{i=1}^3 p_i=1$) are mixtures of bi-separable states with respect to different partitions, and therefore bi-separable. However, as can be easily checked, $\rho$ is entangled with respect to each fixed bipartition, that is, it can not be written as a convex combination of bi-separable states with respect to a fixed bipartition.

Before we formulate our separability criteria, an introduction of
notations that will be involved in the subsequent
sections of our article is necessary. Let $P_i$
 be the operator swapping the two copies of $\mathcal{H}_i$ in $\mathcal{H}^{\otimes 2}=(\mathcal{H}_1\otimes \mathcal{H}_2\otimes\cdots\otimes\mathcal{H}_n)^{\otimes 2}$ , i.e. it performs a permutation on $\mathcal{H}_i^{\otimes 2}$  and leaves all other subsystems unchanged $P_i|x_1\cdots x_{i-1}x_ix_{i+1}\cdots x_n\rangle|y_1\cdots y_{i-1}y_iy_{i+1}\cdots y_n\rangle=|x_1\cdots x_{i-1}y_ix_{i+1}\cdots x_n\rangle|y_1\cdots y_{i-1}x_iy_{i+1}\cdots y_n\rangle$, while $P_i^\dag$  is the adjoint or Hermitian conjugate of the operator  $P_i$.  $P_{tot}$  denotes the operator that performs a simultaneous local permutation on all subsystems in $\mathcal{H}^{\otimes 2}$, that is $P_{tot}=P_1\circ P_2\circ \cdots \circ P_n$. A simple example would be  $P_{tot}|x_1x_2 \cdots x_n\rangle|y_1y_2\cdots y_n\rangle=|y_1y_2\cdots y_n\rangle|x_1x_2 \cdots x_n\rangle$.

\section{The $k$-separability criteria for $n$-partite quantum states }

Now, we state our main results:

\textbf{Theorem 1} ~ Suppose that $\rho$ is an $n$-partite density
matrix acting on Hilbert space $\mathcal{H}=\mathcal{H}_1\otimes
\mathcal{H}_2\otimes\cdots\otimes\mathcal{H}_n$ with dim$\mathcal{H}_l=d_l\geq 2$. Let $|\Phi_{ij}\rangle=|\phi_i\rangle|\phi_j\rangle$, where $|\phi_i\rangle=|x_1x_2\cdots x_{i-1}\tilde{x}_ix_{i+1}\cdots x_n\rangle$ are fully separable states of $\mathcal{H}$.   If $\rho$ is $k$-separable, then
\begin{equation}\label{k-separable-phi}
\begin{array}{cl}
 & \sum\limits_{i\neq j}\sqrt{\langle\Phi_{ij}|\rho^{\otimes
2}P_{tot}|\Phi_{ij}\rangle} \\
\leq & \sum\limits_{i\neq
j}\sqrt{\langle\Phi_{ij}|P_i^\dag\rho^{\otimes
2}P_i|\Phi_{ij}\rangle} \\
& +(n-k)\sum\limits_{i}\sqrt{\langle\Phi_{ii}|P_i^\dag\rho^{\otimes
2}P_i|\Phi_{ii}\rangle}.
 \end{array}
\end{equation}
Of course, $\rho$ is a $k$-nonseparable $n$-partite state if it violates
the above ineq.(\ref{k-separable-phi}).

\textbf{Proof.} ~ The proof is similar to that of Theorem 1 in Ref.\cite{GaoHong2010}.   To establish the validity of ineq.(\ref{k-separable-phi}) for all $k$-separable states $\rho$, let us
first verify that this is true for any $k$-separable pure state
$\rho$.

Suppose that $\rho=|\psi\rangle\langle\psi|$ is a
 $k$-separable pure state under  a $k$-partition $A_1|A_2|\cdots|A_k$, and
$
   |\psi\rangle= |\psi_1\rangle_{A_1}\cdots|\psi_k\rangle_{A_k}.
$

 By calculation, one has
\begin{equation}\label{iandj}
\begin{array}{rl}
   & \sqrt{\langle\Phi_{ij}|\rho^{\otimes
2}P_{tot}|\Phi_{ij}\rangle}
 =
 \sqrt{\langle\phi_i|\rho|\phi_i\rangle\langle\phi_j|\rho|\phi_j\rangle}
 \\
 \leq & \frac{\sqrt{\langle\Phi_{ii}|P_i^\dag\rho^{\otimes
2}P_i|\Phi_{ii}\rangle}+\sqrt{\langle\Phi_{jj}|P_j^\dag\rho^{\otimes
2}P_j|\Phi_{jj}\rangle}}{2}
\end{array}
\end{equation}
in case of $i,j$ in same part, and
\begin{equation}\label{iorj}
\begin{array}{rl}
 \sqrt{\langle\Phi_{ij}|\rho^{\otimes
2}P_{tot}|\Phi_{ij}\rangle}
 = &
 \sqrt{\langle\phi|\rho|\phi\rangle\langle\phi_{ij}|\rho|\phi_{ij}\rangle}
 \\
  = & \sqrt{\langle\Phi_{ij}|P_i^\dag\rho^{\otimes
2}P_i|\Phi_{ij}\rangle}
\end{array}
\end{equation}
in case of $i,j$ in different parts ( $i\in A_l, j\in A_{l'}$ with
$l\neq l'$). Here $|\phi\rangle=\otimes_{i=1}^n|x_i\rangle=|x_1x_2\cdots x_n\rangle$ and  $|\phi_{ij}\rangle=|x_1\cdots x_{i-1}\tilde{x}_ix_{i+1}\cdots x_{j-1}\tilde{x}_jx_{j+1}\cdots  x_n\rangle$ are  fully separable states of $\mathcal{H}$.
 Combining
(\ref{iandj}) and (\ref{iorj}) gives that {\small
\begin{equation}\label{}
\begin{array}{rl}
   ~~ & \sum\limits_{i\neq j}\sqrt{\langle\Phi_{ij}|\rho^{\otimes
2}P_{tot}|\Phi_{ij}\rangle} \\
  = & \sum\limits_{i\in A_l,j\in A_{l'}, l\neq{l'}\atop
 l,l'\in\{1, 2, \cdots, k\} } \sqrt{\langle\Phi_{ij}|\rho^{\otimes
2}P_{tot}|\Phi_{ij}\rangle}\\
& +\sum\limits_{i,j\in A_l, i\neq j \atop l\in\{1, 2, \cdots, k\} }\sqrt{\langle\Phi_{ij}|\rho^{\otimes
2}P_{tot}|\Phi_{ij}\rangle} \\
\leq &  \sum\limits_{i\in A_l,j\in A_{l'}, l\neq{l'}
\atop l,l'\in\{1, 2, \cdots, k\} } \sqrt{\langle\Phi_{ij}|P_i^\dag\rho^{\otimes
2}P_i|\Phi_{ij}\rangle}\\
& +\sum\limits_{i,j\in A_l, i\neq j \atop l\in\{1, 2, \cdots, k\} }(\frac{\sqrt{\langle\Phi_{ii}|P_i^\dag\rho^{\otimes
2}P_i|\Phi_{ii}\rangle}+\sqrt{\langle\Phi_{jj}|P_j^\dag\rho^{\otimes
2}P_j|\Phi_{jj}\rangle}}{2}) \\
\leq &  \sum\limits_{i\neq
j}\sqrt{\langle\Phi_{ij}|P_i^\dag\rho^{\otimes
2}P_i|\Phi_{ij}\rangle}\\
& +(n-k)\sum\limits_{i}\sqrt{\langle\Phi_{ii}|P_i^\dag\rho^{\otimes
2}P_i|\Phi_{ii}\rangle}.
\end{array}
\end{equation} }
Here we have used ineqs. (\ref{iandj}) and (\ref{iorj}), and the number $m_l$ of elements in $A_l$  is at most $n-k+1$ because of $m_l\geq 1$.
Thus, ineq.(\ref{k-separable-phi}) is satisfied by all $k$-separable
$n$-partite pure states.

It remains to show that ineq.(\ref{k-separable-phi})  holds if $\rho$ is
a $k$-separable $n$-partite mixed state. Indeed, the generalization
of ineq.(\ref{k-separable-phi}) to mixed states is a direct
consequence of the convexity of its left hand side  and  the
concavity of its right hand side, which we can see in the following.

Suppose that
$
 \rho=\sum \limits_mp_m\rho_m=\sum
\limits_mp_m|\psi_m\rangle\langle\psi_m|
$
is a $k$-separable $n$-partite mixed state, where
$\rho_m=|\psi_m\rangle\langle\psi_m|$ is $k$-separable. Then,  by
Cauchy-Schwarz inequality $(\sum\limits_{k=1}^mx_ky_k)^2\leq(\sum\limits_{k=1}^mx_k^2)(\sum\limits_{k=1}^my_k^2)$,
we have {\footnotesize
\begin{equation}\label{}
\begin{array}{rl}
 & \sum\limits_{i\neq j}\sqrt{\langle\Phi_{ij}|\rho^{\otimes
2}P_{tot}|\Phi_{ij}\rangle}
\leq  \sum\limits_mp_m\sum\limits_{i\neq
j}\sqrt{\langle\Phi_{ij}|\rho_m^{\otimes 2}
P_{tot}|\Phi_{ij}\rangle} \\
\leq &  \sum\limits_mp_m\bigg(\sum\limits_{i\neq
j}\sqrt{\langle\Phi_{ij}|P_i^\dag\rho_m^{\otimes
2}P_i|\Phi_{ij}\rangle}\\
& +(n-k)\sum\limits_{i}\sqrt{\langle\Phi_{ii}|P_i^\dag\rho_m^{\otimes 2}P_i|\Phi_{ii}\rangle}\bigg) \\
= &  \sum\limits_{i\neq
j}\sum\limits_m\sqrt{\langle\phi|p_m\rho_m|\phi\rangle}\sqrt{\langle\phi_{ij}|p_m\rho_m|\phi_{ij}\rangle}\\
& +(n-k)\sum\limits_{i}\sum\limits_mp_m\langle\phi_i|\rho_m|\phi_i\rangle \\
\leq  &  \sum\limits_{i\neq
j}\sqrt{\sum\limits_m\langle\phi|p_m\rho_m|\phi\rangle\sum\limits_m\langle\phi_{ij}|p_m\rho_m|\phi_{ij}\rangle}
\\
& +(n-k)\sum\limits_{i}\langle\phi_i|\rho|\phi_i\rangle \\
= & \sum\limits_{i\neq j}\sqrt{\langle\Phi_{ij}|P_i^\dag\rho^{\otimes
2}P_i|\Phi_{ij}\rangle} \\
& +(n-k)\sum\limits_{i}\sqrt{\langle\Phi_{ii}|P_i^\dag\rho^{\otimes
2}P_i|\Phi_{ii}\rangle},
\end{array}
\end{equation}  }
as desired. This completes the proof.

The special case of above Theorem 1 is:

\textbf{Corollary} ~  Let $\rho$ be an  $n$-partite density
matrix acting on Hilbert space $\mathcal{H}=\mathcal{H}_1\otimes
\mathcal{H}_2\otimes\cdots\otimes\mathcal{H}_n$, and $|\Phi_{ij}\rangle=|\phi_i\rangle|\phi_j\rangle$ with
$|\phi_i\rangle=|x\cdots xyx\cdots x\rangle\in \mathcal{H}$, where the local
state of $\mathcal{H}_l$ is $|x\rangle$ for $l\not=i$ and
$|y\rangle$ for $l=i$. Then {\small
\begin{equation}\label{k-separable-phi(special)}
\begin{array}{cl}
& \sum\limits_{i\neq j}\sqrt{\langle\Phi_{ij}|\rho^{\otimes
2}P_{tot}|\Phi_{ij}\rangle}\\
\leq & \sum\limits_{i\neq
j}\sqrt{\langle\Phi_{ij}|P_i^+\rho^{\otimes
2}P_i|\Phi_{ij}\rangle} \\
 & +(n-k)\sum\limits_{i}\sqrt{\langle\Phi_{ii}|P_i^+\rho^{\otimes
2}P_i|\Phi_{ii}\rangle}.
\end{array}
\end{equation} }
 If an $n$-partite state $\rho$ does not satisfy the above ineq.(\ref{k-separable-phi(special)}), then $\rho$ is  not $k$-separable ($k$-nonseparable).

\textbf{Theorem 2.} Every fully separable $n$-partite state $\rho$ satisfies
\begin{equation}
\sqrt{\langle\Phi_{ij}|\rho^{\otimes2}P_{tot}|\Phi_{ij}\rangle}\leq\sqrt{\langle
\Phi_{ij}|P_{i}^{\dagger}\rho^{\otimes2}P_{i}|\Phi_{ij}\rangle}
\label{FullySeparable}\end{equation}
for fully separable states defined as
$
 |\Phi_{ij}\rangle =|\phi_{i}\rangle
|\phi_{j}\rangle, $
 with
$|\phi_{i}\rangle =|x_1\cdots x_{i-1}\tilde{x}_ix_{i+1}\cdots x_n\rangle
\in \mathcal{H}=\mathcal{H}_{1}\otimes\mathcal{H}_{2}\otimes\cdots\otimes\mathcal{H}_{n}$,
where the $i$-th local state is $|\tilde{x}_i\rangle $ and the
others are $|x_k\rangle $ ($k\neq i$).
 These are $\frac{1}{2}n(n-1)$
inequalities, and violation of any one of them implies
nonseparability.

\textbf{Proof.} ~ Note that the left-hand side of ineq.(\ref{FullySeparable}) minus
the right-hand side of (\ref{FullySeparable}) is a convex function of the matrix $\rho$
entries (since the left-hand side is convex and the right-hand side is
concave). Consequently, it suffices to prove
the validity for fully separable pure states, and the validity for
mixed states is guaranteed.

Suppose that $\rho$ is a fully separable $n$-partite pure state, then one has
\begin{equation}
\sqrt{\langle
\Phi_{ij}|\rho^{\otimes2}P_{tot}|\Phi_{ij}\rangle
}=|\langle \phi_{i}|\rho|\phi_{j}\rangle|,
\end{equation}
 and
\begin{equation}\label{2}
\begin{array}{rl}
& \sqrt{\langle\Phi_{ij}|P_i^{\dagger}\rho^{\otimes2}P_i|\Phi_{ij}\rangle }  \\
= & \sqrt{\langle \phi|\rho|\phi\rangle
\langle \phi_{ij}|\rho|\phi_{ij}\rangle}=|\langle
\phi_{i}|\rho|\phi_{j}\rangle|.
\end{array}
\end{equation}
Here $|\phi\rangle=\otimes_{i=1}^n|x_i\rangle=|x_1x_2\cdots x_n\rangle$ and  $|\phi_{ij}\rangle=|x_1\cdots x_{i-1}\tilde{x}_ix_{i+1}\cdots x_{j-1}\tilde{x}_jx_{j+1}\cdots  x_n\rangle$ are  fully separable states of Hilbert space $\mathcal{H}$.
The combination of above two equalities gives that (\ref{FullySeparable}) holds with equality if  $\rho$ is a fully separable $n$-partite pure state.

 The special case of Theorem 1 with $k=n$  can be given exactly by taking the sum of all the inequalities in Theorem 2, indicating that the latter can serve as a tighter criterion for $n$-nonseparability.  Theorem 2 is not the particular case of Theorem 1 with $k=n$.

Our $k$-separability criteria can be used to distinguish the different classes of multipartite  inseparable states and can detect many important multipartite entangled states  such as GHZ state, W state, and anti W state efficiently. They can be used for detecting not only genuine $n$-partite entangled mixed states ($k=2$) but also $k$-nonseparable mixed multipartite states (not $k$-separable states) ($k=3,4,\cdots,n$). Moreover, it indeed detects $k$-nonseparable mixed multipartite states [for $n$-qubit states such as  W state mixed with white noise, the mixture of the identity matrix, the W state and the anti-W state, and the mixture of the GHZ state, the W state and the identity matrix ] which beyond all previously studied criteria. Theorem 1 in Ref.\cite{GaoHong2010} is the special case of our Corollary above when $k=2$  and can only be used to identify genuine $n$-partite entangled mixed states ($k=2$).

\section{Examples}

In this section, we illustrate our main result with some explicit examples.
It should be pointed out that our criteria are suitable for any $n$-partite states. For simplicity, we give the following examples to show the detecting ability of our criteria.

\textit{Example 1} ~ Consider the family of $n$-qubit states
\begin{equation}\label{W}
\rho^{(G-W_n)}=\alpha|\mathrm{GHZ}_n\rangle\langle
\mathrm{GHZ}_n|+\beta|W_n\rangle\langle
W_n|+\frac{1-\alpha-\beta}{2^n}\mathbf{I},
\end{equation}
the mixture of the GHZ state and the W state, dampened by isotropic
noise. Here $|\mathrm{GHZ}_n\rangle=\frac{1}{\sqrt{2}}(|00\cdots 0\rangle+|11\cdots 1\rangle)$
and
$
|W_n\rangle=\frac{1}{\sqrt{n}}(|0\cdots 01\rangle+|0\cdots
10\rangle+\cdots+|1\cdots 00\rangle)
$
are the $n$-qubit GHZ state and W state, respectively.

For the family $\rho^{(G-W_n)}$, our criteria can detect $k$-nonseparable states that had not been identified so far.

The detection parameter spaces of our $k$-separability criteria (ineq.(\ref{k-separable-phi}) ) and that in Ref.\cite{A.Gabriel2010}  for $n=4, k=3$ are illustrated in  Fig.1.

\begin{figure}
\begin{center}
{\includegraphics[scale=0.4]{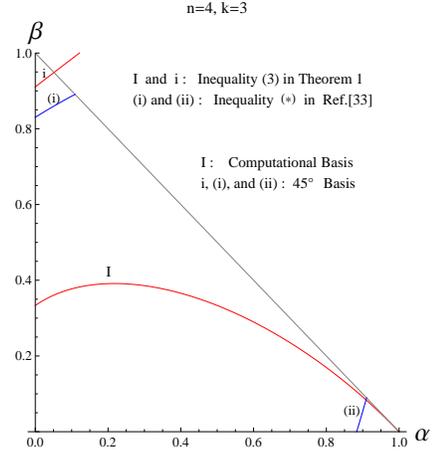}} \caption[Illustration of the
detection parameters of ineq.(\ref{k-separable-phi}) in
Theorem 1 and ineq.($\ast$) in \cite{A.Gabriel2010}]{(Color
online). Here the detection quality of ineq.(\ref{k-separable-phi}) in
Theorem 1 and the  ineq.($\ast$) in
\cite{A.Gabriel2010} is shown for the state
$\rho^{(G-W_n)}=\alpha|\mathrm{GHZ}_n\rangle\langle
\mathrm{GHZ}_n|+\beta|\mathrm{W}_n\rangle\langle
\mathrm{W}_n|+\frac{1-\alpha-\beta}{2^n}\mathbf{I}$, $n=4, k=3$. The lines
I (red line) and i (red line) represent the thresholds of the detection for 3-nonseparable states identified by ineq.(\ref{k-separable-phi}) in
Theorem 1 in the computational basis and $45^\circ$ basis, respectively.
The lines (i) (blue line) and (ii) (blue line) represent the thresholds of the detection for 3-nonseparable states identified by the ineq.($\ast$) in
\cite{A.Gabriel2010} in the $45^\circ$ basis. There is no 3-nonseparable states identified by the ineq.($\ast$) in \cite{A.Gabriel2010} in the computational basis.
The states in the region above the line I identify entanglement ($k=3$) detected by ineq.(\ref{k-separable-phi}) in Theorem 1. The area encircled by the curve I,  the $\beta$ axis, the line (i), the line $\alpha+\beta=1$, and the line (ii) contains 3-nonseparable states detected only by our ineq.(\ref{k-separable-phi}) in
Theorem 1. }
\end{center}
\end{figure}

\textit{Example 2} ~ Consider the $n$-qubit state, W state mixed
with white noise,
\begin{equation}\label{W}
\rho^{(W_n)}(\beta)=\beta|W_n\rangle\langle
W_n|+\frac{1-\beta}{2^n}\mathbf{I}.
\end{equation}
By our Theorem with $|\phi_i\rangle=|x_1\cdots x_{i-1}x_ix_{i+1}\cdots x_n\rangle=|0\cdots 010\cdots 0\rangle\in \mathcal{H}$,  one can derive that  if
\begin{equation}\label{}
1\geq \beta>\frac{n(2n-k-1)}{2^n(k-1)+n(2n-k-1)},
\end{equation}
then $\rho^{(W_n)}(\beta)$ is not $k$-separable. In the case $k=2$,
ineq.(\ref{k-separable-phi})  detects W state mixed
with white noise, $\rho^{(W_n)}(\beta)$, for $1\geq
\beta>\frac{n(2n-3)}{n(2n-3)+2^n}$ as genuinely $n$-partite entangled,
whereas ineq.(III) of Ref.\cite{MarcusPRL2010} detects it
for $1\geq \beta>\frac{n^2(n-2)}{n^2(n-2)+2^n}$.

\textit{Example 3} ~ Consider the $n$-qubit state family  given by a mixture of the identity matrix, the W state and the anti-W state
\begin{equation}\label{n-qubit}
\rho^{(W-\tilde{W}_n)}=\frac{1-a-b}{2^n}I_{2^n}+a|W_n\rangle\langle W_n|+b|\tilde{W}_n\rangle\langle \tilde{W}_n|,
\end{equation}
where $|W_n\rangle$ is $n$-qubit W state and $|\tilde{W}_n\rangle=\frac{1}{\sqrt{n}}(|11\cdots110\rangle+|11\cdots101\rangle+\cdots+|01\cdots111\rangle)$. For
this family, our criteria can detect $k$-nonseparable states which have previously not been identified.

 Let {\small $|\Phi\rangle\in\{|0\rangle^{\otimes n}|0\rangle^{\otimes n}, |1\rangle^{\otimes n}|1\rangle^{\otimes n}, |0\rangle^{\otimes n}|1\rangle^{\otimes n}, |1\rangle^{\otimes n}|0\rangle^{\otimes n}, \\ (\frac{|0\rangle+|1\rangle}{\sqrt{2}})^{\otimes
n}(\frac{|0\rangle-|1\rangle}{\sqrt{2}})^{\otimes n}, (\frac{|0\rangle-|1\rangle}{\sqrt{2}})^{\otimes
n}(\frac{|0\rangle+|1\rangle}{\sqrt{2}})^{\otimes n}\}$ }. When $n\geq 4$, ineq.$(\ast)$ in Ref.\cite{A.Gabriel2010}  can not detect genuine $n$-partite entanglement for the family $\rho^{(W-\tilde{W}_n)}$. It can not detect both 3-nonseparable and 4-nonseparable states for the family $\rho^{(W-\tilde{W}_5)}$.

Fig.2 illustrates the entanglement area detected by our ineq.(\ref{k-separable-phi}), ineq.$(\ast)$ in Ref.\cite{A.Gabriel2010} and Proposition 2 in Ref.\cite{GHGcovariancematrix10} for $\rho^{(W-\tilde{W}_3)}$, respectively. The area detected by our ineq.(\ref{k-separable-phi}) is the  largest.

 The detection parameter spaces of our ineq.(\ref{k-separable-phi}) and ineq.$(\ast)$ in Ref.\cite{A.Gabriel2010} for $\rho^{(W-\tilde{W}_4)}$  and $\rho^{(W-\tilde{W}_5)}$  are
illustrated in Fig.3 and Fig.4, respectively. The space detected by the former is visibly larger.

\begin{figure}
    \begin{center}
    {\includegraphics[scale=0.56]{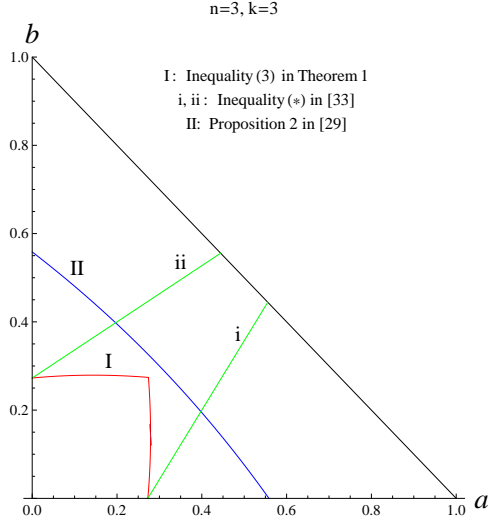}} \caption[Illustration]{(Color
online). Illustration of the detection with our ineq.(\ref{k-separable-phi}), ineq.$(\ast)$ in Ref.\cite{A.Gabriel2010} and Proposition 2 in Ref.\cite{GHGcovariancematrix10} for $\rho^{(W-\tilde{W}_3)}=\frac{1-a-b}{8}I_{8}+a|W_3\rangle\langle W_3|+b|\tilde{W}_3\rangle\langle \tilde{W}_3|$ .  Here the red line I represents the threshold given by ineq.(\ref{k-separable-phi}) in Theorem 1 such that the region above it identifies 3-nonseparable (not fully separable) states. The region  above the blue line II corresponds to  entangled states (not fully separable) detected by Proposition 2 in Ref.\cite{GHGcovariancematrix10}.  The green lines  i and ii represent the thresholds given by ineq.$(\ast)$ in Ref.\cite{A.Gabriel2010} such that   the area enclosed by the line ii (green), the $b$ axis, and line $a+b=1$, and the area  enclosed by the line i (green), line $a+b=1$, and the $a$ axis are not 3-separable (3-nonseparable). So  the area enclosed by the red line I,  green line ii, blue line II, and green line i are not fully separable states detected only by  our ineq.(\ref{k-separable-phi}) in Theorem 1. }
 \end{center}
  \end{figure}

\begin{figure}
    \begin{center}
    {\includegraphics[scale=0.56]{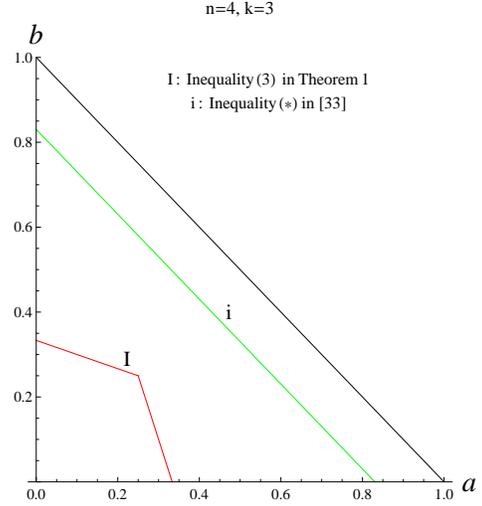}} \caption[ Illustration of the
detection parameter spaces of our ineq.(\ref{k-separable-phi}) in
Theorem 1 and ineq.$(\ast)$ in Ref.\cite{A.Gabriel2010} ]{(Color
online). The detection quality  of our ineq.(\ref{k-separable-phi}) in
Theorem 1 and ineq.$(\ast)$ in Ref.\cite{A.Gabriel2010} is shown for the state
 $\rho^{(W-\tilde{W_4})}=\frac{1-a-b}{16}\textrm{I}_{16}+a|W_4\rangle\langle W_4|+b|\tilde{W_4}\rangle\langle \tilde{W_4}|$, $k=3$.  The red line I represents the threshold given by our ineq.(\ref{k-separable-phi}) such that the region above it corresponds to 3-nonseparable states. The region above the green line i are not 3-separable states detected by ineq.$(\ast)$ in Ref.\cite{A.Gabriel2010}. The area enclosed by the red line I, the $a$ axis, the green line i, and the $b$ axis is the entanglement (3-nonseparable) detected only by  our ineq.(\ref{k-separable-phi}) in Theorem 1.}
   \end{center}
    \end{figure}

\begin{figure}
    \begin{center}
    {\includegraphics[scale=0.56]{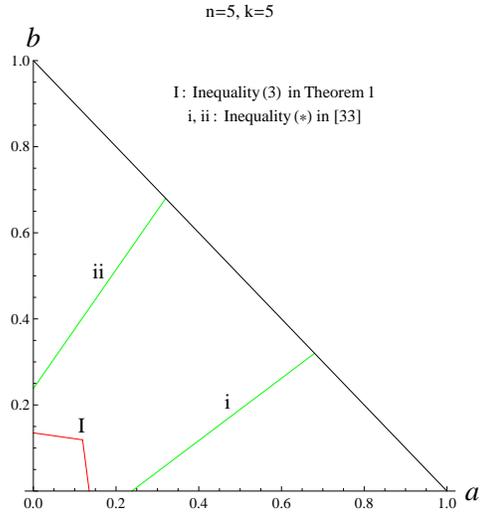}} \caption[Illustration]{(Color
online). Illustration of the detection with our ineq.(\ref{k-separable-phi}) in Theorem 1 and ineq.$(\ast)$ in \cite{A.Gabriel2010} for $\rho^{(W-\tilde{W_5})}=\frac{1-a-b}{32}\textrm{I}_{32}+a|W_5\rangle\langle W_5|+b|\tilde{W_5}\rangle\langle \tilde{W_5}|$, $k=5$.
The region  above the red line I  are not fully separable states  detected by our ineq.(\ref{k-separable-phi}) in Theorem 1. Both the area enclosed by
 the green line i, the $a$ axis, and the lines $a+b=1$, and the area enclosed by the green line ii, the line $a+b=1$, and the $b$ axis are  not fully separable given by ineq.$(\ast)$ in \cite{A.Gabriel2010}.
 The area enclosed by the red line I, the $a$ axis, the green line i, the line $a+b=1$, the green line ii, and  the $b$ axis are the entanglement (not full separable) detected only by our ineq.(\ref{k-separable-phi}).}
    \end{center}
    \end{figure}

\section{Experimental implementation}

In this section, we show that our criteria can be easily implemented in experiment without quantum state tomography and then we give the local observables required to implement our criteria. A local observable  is an observable
such as  $L=B_1\otimes B_2\otimes\cdots \otimes B_n$, $L=B_1 B_2\cdots B_n$ for short, where $B_l$ denotes
observable on subsystem $l$, $l=1,2,\ldots, n$. Thus, local observables can be measured locally.

Each term on the left hand side  and  on the right hand side of ineq.(\ref{k-separable-phi}) can be determined
by measuring two observables and a single observable, respectively.
These observables determining ineq.(\ref{k-separable-phi}) can
be implemented by means of local observables. We give the local observables required to implement our criteria using the same method as \cite{GaoHong2010}.
The observables associated with each term (diagonal matrix elements)
of the right hand side in ineq.(\ref{k-separable-phi}) can be
implemented by means of local observables, which can be seen from
the following expressions $|\phi\rangle\langle\phi|=\otimes_{l=1}^n T_l$, $|\phi_{ij}\rangle\langle\phi_{ij}|=T_1\cdots  T_{i-1}Q_i T_{i+1}\cdots  T_{j-1} Q_j T_{j+1}\cdots T_n$, and
$|\phi_i\rangle\langle\phi_i|=T_1\cdots  T_{i-1}Q_i T_{i+1}\cdots  T_n$, where $T_l=|x_l\rangle\langle x_l|$ and
$Q_i=|\tilde{x}_i\rangle\langle \tilde{x}_i|$. Thus, determining one diagonal matrix
element requires only a single local observable.

Each term $\sqrt{\langle\Phi_{ij}|\rho^{\otimes
2}P_{tot}|\Phi_{ij}\rangle}=|\langle\phi_i|\rho|\phi_j\rangle|$ of the right hand side in ineq.(\ref{k-separable-phi})  can be determined
by measuring two observables
$O_{ij}$ and $\tilde{O}_{ij}$, since  $\langle
O_{ij}\rangle=2\mathrm{Re}\langle\phi_i|\rho|\phi_j\rangle$ and
$\langle
\tilde{O}_{ij}\rangle=-2\mathrm{Im}\langle\phi_i|\rho|\phi_j\rangle$.
Here
$O_{ij}=|\phi_i\rangle\langle\phi_j|+|\phi_j\rangle\langle\phi_i|$
and
$\tilde{O}_{ij}=-\mathrm{i}|\phi_i\rangle\langle\phi_j|+\mathrm{i}|\phi_j\rangle\langle\phi_i|$.
Without loss of generality, let $i<j$. From  {\small
\begin{equation}\label{}
\begin{array}{rl}
  O_{ij}= & \frac{1}{2} T_1\cdots  T_{i-1} M_i T_{i+1}\cdots  T_{j-1}
M_j T_{j+1}\cdots  T_n
\\
& + \frac{1}{2} T_1\cdots  T_{i-1} \tilde{M}_i T_{i+1}\cdots  T_{j-1}
\tilde{M}_j T_{j+1}\cdots  T_n,
\end{array}
\end{equation}
\begin{equation}\label{}
\begin{array}{rl}
  \tilde{O}_{ij} =& \frac{1}{2} T_1\cdots  T_{i-1} M_i T_{i+1}\cdots  T_{j-1}
\tilde{M}_j T_{j+1}\cdots  T_n
\\
& - \frac{1}{2} T_1\cdots  T_{i-1} \tilde{M}_i T_{i+1}\cdots  T_{j-1}
M_j T_{j+1}\cdots  T_n,
\end{array}
\end{equation}  }
where $M_i=|\tilde{x}_i\rangle\langle x_i|+|x_i\rangle\langle \tilde{x}_i|$,
$\tilde{M}_i=\mathrm{i}|\tilde{x}_i\rangle\langle x_i|-\mathrm{i}|x_i\rangle\langle
\tilde{x}_i|$, one can determine the left hand side in ineq.(\ref{k-separable-phi}) by $2(n^2-n)$ local observables.

Therefore in total at most $\frac{5(n^2-n)}{2}+n+1$
 local observables are needed
to test our separability criteria ineq.(\ref{k-separable-phi}).  For any unknown  $n$-partite mixed states, experimental detection of multipartite $k$-inseparability using our criteria require only  $\mathcal{O}(n^2)$ local measurements, which is much less than  using the criterion \cite{A.Gabriel2010}   and quantum state tomography that would require $\mathcal{O}(2^n)$ and $(d_1^2-1)(d_2^2-1)\cdots (d_n^2-1)$) local  measurements, respectively.   Thus, our $k$-separability criteria can be used easily for experimental detection of multipartite entanglement.

Of course, entanglement witness can be used for experimental detection of entanglement. It is enough to measure only one observable - entanglement witness - in order to detect entanglement in a given state. However, it can not easily be implemented in experiment if it is not decomposed into operators that can be measured locally. Therefore, for the experimental implementation it is necessary to decompose the witness into operators that can be measured locally. In section 6 of Ref.\cite{GuhneToth2009}, G\"{u}hne and T\'{o}th pointed out: To obtain a good local decomposition requires often some effort, especially proving that a given decomposition is optimal, is often very difficult. For any pure state there exists a witness that requires $2n-1$ measurements, but the robustness to noise may be small. Furthermore, there exist observables, for which the local decomposition requires $\frac{2\bullet 3^{n-1}}{n+1}$ local measurements, which means that a local measurement of these observables requires nearly the same effort as state tomography.

\section{Conclusion}

In this paper, we present  efficient $k$-separability criteria to
identify $k$-nonseparable multipartite mixed states in arbitrary
dimensional quantum systems. The resulting criteria are easily
computable from the density matrix, and no optimization or
eigenvalue computation is needed. Our criteria can be used to distinguish the different classes of multipartite  inseparable states and can detect many important multipartite entangled states  such as GHZ states, W states,  anti W states, and mixtures thereof efficiently. They can be used for detecting not only genuine $n$-partite entangled mixed states ($k=2$) but also $k$-nonseparable mixed multipartite states (not $k$-separable states) ($k=3,4,\cdots,n$). In addition, our criteria detect multipartite entanglement  that had not been identified so far and can be used in today¡¯s experiments without the need for quantum state
tomography: only $\mathcal{O}(n^2)$ local measurement settings are needed.

\acknowledgments
This work was supported by the National Natural Science Foundation
of China under Grant No: 11371005, Hebei Natural Science Foundation
of China under Grant Nos: A2012205013.

\end{document}